\documentclass[	
						twocolumn,%
						a4paper,
						showkeys,
					]{revtex4}

\usepackage[frenchb,english]{babel}
\usepackage[T1]{fontenc}
\usepackage{upgreek}
\usepackage{textcomp} 				
\usepackage[dvips]{graphicx}		
\usepackage{color}					
\usepackage[nice]{nicefrac} 		
\usepackage{amssymb,amsmath}		
\usepackage{array}					
\usepackage{ulem}						

\newcommand{\etal}{\textit{et al.}}
\newcommand{\eps}{\varepsilon}

\begin{document}

\title{High-pressure investigations of CaTiO$_3$ up to 60 GPa using x-ray diffraction and Raman spectroscopy}
\date{\today}
\author{Mael Guennou}
\author{Pierre Bouvier}
\author{Benjamin Krikler}
\author{Jens Kreisel}
\affiliation{Laboratoire des Mat\'eriaux et du G\'enie Physique, CNRS UMR 5628, Grenoble Institute of Technology, MINATEC,
3 parvis Louis N\'eel, 38016 Grenoble, France}
\author{Raphaël Haumont}
\affiliation{Laboratoire de Physico-Chimie de l'\'Etat Solide, ICMMO, CNRS UMR 8182, Universit\'e Paris XI, 91405 Orsay, France}
\author{Gaston Garbarino}
\affiliation{European Synchrotron Radiation Facility (ESRF), BP 220, 6 Rue Jules Horowitz, 38043 Grenoble Cedex, France}

\begin{abstract}
In this work, we investigate calcium titanate (CaTiO$_3$ - CTO) using X-ray diffraction and Raman spectroscopy up to 60 and 55 GPa respectively. Both experiments show that the orthorhombic $Pnma$ structure remains stable up to the highest pressures measured, in contradiction to ab-initio predictions. A fit of the compression data with a second-order Birch-Murnaghan equation of state yields a bulk modulus $K_0$ of 181.0(6) GPa. The orthorhombic distortion is found to increase slightly with pressure, in agreement with previous experiments at lower pressures and the general rules for the evolution of perovskites under pressure. High-pressure polarized Raman spectra also enable us to clarify the Raman mode assignment of CTO and identify the modes corresponding to rigid rotation of the octahedra, A-cation shifts and Ti-O bond stretching. The Raman signature is then discussed in terms of compression mechanisms.
\end{abstract}


\maketitle

\section{Introduction}

In the past, investigations of $AB$O$_3$ perovskite-type oxides have been a rich source for the understanding of structural mechanisms not only in perovskites but also in oxides and other ionic materials in general. Beyond their fundamental interest, perovskites attract continuing attention as functional oxides, often related to dielectric, ferroelectric and multiferroic properties \cite{Rabe2007,Kreisel2009}. The ideal cubic structure of $AB$O$_3$ perovskites is rather simple, with corner-linked $B$O$_6$ octahedra and the $A$ cations sitting in the space between the octahedra \cite{Glazer1972,Glazer1975,Mitchell2002}. Most perovskites present, however, structural distortions away from this parent cubic structure, which can be driven by external parameters such as temperature, pressure, stress etc. One of the major challenges in perovskite science is to identify these distortions, which are often rather subtle and thus difficult to follow. 

Here we will focus on a class of materials in which the structural distortion is dominated by tilts (rotations) of the $B$O$_6$ octahedra \cite{Glazer1972}. This kind of distortion is ferroelastic and corresponds to anti-ferrodistortive (AFD) instabilities at the zone boundary \cite{Salje1993}. It is well known \cite{Mitchell2002} that such a tilt distortion (the tilt angle) can be driven to a large extent by temperature or pressure as exemplified in the model materials SrTiO$_3$ \cite{Fleury1968,Guennou2010} or LaAlO$_3$ \cite{Scott1969,Bouvier2002,Hayward2005}. It is generally accepted that temperature reduces tilt instabilities, i.e. the tilt angle decreases with increasing temperature. The situation is less straightforward for the effect of high-pressure: based on a pioneering work by Samara \etal{} \cite{Samara1975}, it was for a long time believed that pressure increases systematically AFD tilt instabilities, i.e. the tilt angle increases with increasing pressure. However, more recently, this view has been challenged by a number of counter-examples \cite{Bouvier2002,Angel2005,Tohei2005}. This situation has motivated the search for criteria able to predict this evolution with pressure. Generally speaking, the volume reduction in perovskites can be accomodated by two different mechanisms \cite{Zhao2004,Zhao2006,Angel2005}: bond compression that favours a reduction of the distortion, and tilting of the octahedra that results in an increase of the distortion. The pressure dependence of the distortion is then controlled by the relative compressibilities $\beta$ of the $B$O$_6$ and $A$O$_{12}$ polyhedra (or equivalently of the $A$-O and $B$-O bonds): $\beta_B/\beta_A < 1$ implies an increase of the tilt angles under pressure and vice versa. Experimentally, Andrault and Poirier \cite{Andrault1991} first estimated the individual polyhedron compressibilities using the empirical relation between the volume and the bulk modulus of the polyhedra given by Hazen and Finger \cite{Hazen1982}. More recently Zhao, Ross and Angel \cite{Zhao2004,Zhao2006} have developed a model based on the calculation of the valence bond sums, whose main feature is the prediction that the tilt angle should increase under pressure for $A^{2+}B^{4+}$O$_3$ perovskites but decrease for $A^{3+}B^{3+}$O$_3$ compounds, such as aluminates $A$AlO$_3$, in agreement with experimental data. Moreover, the bond valence sum model provides a way to estimate the angle change under pressure \cite{Zhao2006}.

Having the evidence that octahedra tilts can either increase or decrease with increasing pressure, it is now interesting to ask if there are perovskites in which the tilt angle does not change with pressure, but where the compression is largely dominated by bond compression, and up to what point such structures can remain stable. To the best of our knowledge, the search for such materials has until now received only little attention, with the notable exceptions of CaTiO$_3$ (CTO) and CaGeO$_3$ \cite{Ross1999} whose tilt angles and spontaneous strains show only little variations up to 10 GPa. CTO has been in the past often investigated both at low temperatures for its intriguing dielectric properties \cite{Lemanov1999} and at high pressures and temperatures for its suggested analogy with geologically relevant perovskites such as MgSiO$_3$ \cite{Gillet1993}. Previous investigations at high pressure have been carried out using Raman spectroscopy up to 26 GPa \cite{Gillet1993} and X-ray diffraction in various pressure ranges \cite{Xiong1986,Ross1999,Wu2004}. Although Xiong \etal{} \cite{Xiong1986} have claimed a transition to a hexagonal phase at approximately 10 GPa, this has not been confirmed by either the Raman scattering experiment by Gillet \etal{} \cite{Gillet1993}, or the diffraction experiment by Wu \etal{} \cite{Wu2004}.

In this work we carry out experiments on CTO single crystals at room temperature (RT) by using Raman scattering and X-ray diffraction. The general aim of our study is to follow and interpret the structural distortions up to high pressure, and more specifically: (i) to clarify the presence of a theoretically predicted \cite{Wu2005a} and experimentally disputed \cite{Xiong1986,Gillet1993,Wu2004} pressure-induced phase transition and (ii) to explore the compression mechanisms at very high pressure up to 60 GPa. 

\section{Experimental details}

\begin{figure}[tb]
\begin{center}
\includegraphics[width=0.47\textwidth]{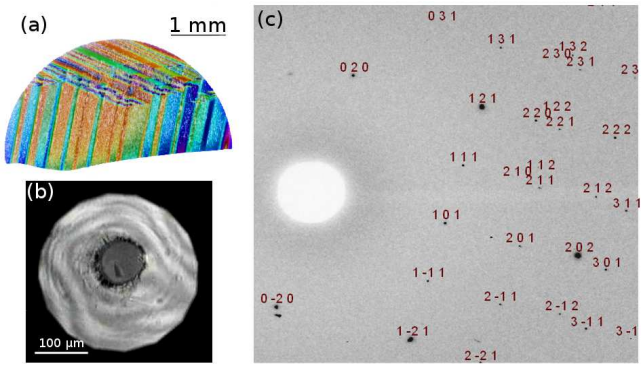}
\caption{(a) Polarized light micrograph of a slice cut from the as-grown crystal, showing large ferroelastic domains. (b) Picture of the pressure chamber at 60 GPa during the X-ray diffraction experiment. (c) Example of a diffraction pattern integrated over the full angular range with peak indexing.}
\label{fig:clicheRX}
\end{center}
\end{figure}

CaTiO$_3$ single crystals were grown by the floating-zone technique, with powders of CaCO$_3$ and TiO$_2$ (99.9\% purity) as starting materials. The as-grown crystals showed a clear domain structure visible under polarized light, with approximately 100 $\upmu$m sized domains (figure \ref{fig:clicheRX}(a)). The samples used for the experiments were polished to a thickness of about 10 $\upmu$m with a lateral extension of 10 to 30 $\upmu$m and did not show any domain structure, neither under polarized light nor in the diffraction patterns. All experiments were performed in diamond-anvil cells (DAC) with various pressure transmitting media, as detailed in the following. The cells have the Boehler-Almax diamond design with a cullet of 300 $\upmu$m. The pressure chamber was sealed by a stainless steel gasket pre-indented to a thickness of about 40 $\upmu$m. We checked by measuring the gasket thickness after the experiment that the pressure chamber was always larger than the crystal in order to rule out any bridging of the diamonds by the sample. A picture of the pressure chamber at 60 GPa is shown in figure \ref{fig:clicheRX}(b).

X-ray diffraction experiments were performed on the ID27 beamline at the ESRF. The beam was monochromatic with a wavelength of 0.3738 \AA{} selected by an iodine K-edge filter and focused to a beam size of about 3 $\upmu$m. The signal was collected in the rotating crystal geometry (-30\degre{}$\le\omega\le$ 30\degre) on a marCCD detector. The crystal has its $c$ axis close to the DAC axis, as can be infered from the indexed pattern given in figure \ref{fig:clicheRX}(c). A precise calibration of the detector parameters was performed with a reference silicon powder. Helium was used as a pressure transmitting medium. The fluorescence of ruby was used as a pressure gauge \cite{Mao1978}. During our experiments, we have carefully followed the splitting between the two fluorescence lines R1 and R2 of ruby: the splitting was approximately 1.40 nm at ambient conditions and drifted slightly under pressure up to a maximum of 1.50 nm at the highest pressure measured.

Raman scattering experiments were performed with neon as a pressure transmitting medium. The spectra were recorded on a Labram spectrometer with a low frequency cutoff at 100 cm$^{-1}$. The exciting laser line was the Argon at 514.5 nm. The laser power was kept at 10 mW on the DAC to avoid heating of the sample. 

\section{X-ray diffraction}

At ambient conditions, CTO is orthorhombic with space group $Pnma$ \cite{Sasaki1987}. The oxygen octahedra are tilted along the three directions of the cubic perovskite unit cell, with a tilt system $a^-b^+a^-$ using Glazer's notation \cite{Mitchell2002,Glazer1972,Glazer1975}. The presence of twins is rather common in $Pnma$ perovskites in general and may complicate considerably the analysis of the diffraction pattern. In our case however, the analysis of the superstructure reflections corresponding to the tilts revealed that our crystal was a single domain crystal, with only very weak reflections arising from a possible twin. The diffraction spots could therefore be indexed unambiguously in the $Pnma$ space group. The diffraction pattern did not show any major change over the entire pressure range investigated and the same indexation was used throughout the experiment. The lattice parameters were determined by a least squares fit to the positions of the observed peaks performed with the program UnitCell \cite{Holland1997}. For each pressure, about 100 peak positions were refined, covering $d$-spacings down to 1.067 \AA. The lattice constants are reported in table \ref{tab:latticeconstants}. 

\begin{table}[tb]
\begin{center}
\begin{tabular}{r l l l}
\hline\hline
P (GPa) 		& $a_{\mathrm{pc}}$ (\AA) & $b_{\mathrm{pc}}$ (\AA) & $c_{\mathrm{pc}}$ (\AA) \\\hline
0.19(20)		& 3.8448(10)&	3.8201(12)& 3.7990(32)\\
0.81(11)		& 3.8402(10)&	3.8161(12)&	3.7945(34)\\
2.19(17)		& 3.8311(13)&	3.8049(15)&	3.7842(39)\\
3.14(11)		& 3.8241(8) &	3.8000(10)&	3.7785(27)\\
4.77(26)		& 3.8136(10)&	3.7898(12)&	3.7668(31)\\
6.20(23)		& 3.8047(9) &	3.7793(10)&	3.7576(28)\\
7.74(21)		& 3.7954(10)&	3.7705(12)&	3.7490(33)\\
9.38(9)		& 3.7855(10)&	3.7612(11)&	3.7391(31)\\
10.37(12)	& 3.7803(10)&	3.7568(11)&	3.7330(33)\\
12.08(22)	& 3.7710(8) &	3.7469(10)&	3.7230(28)\\
13.94(21)	& 3.7612(8) &  3.7362(9)&	3.7133(27)\\
16.14(19)	& 3.7498(8) &	3.7253(10)&	3.7025(28)\\
18.19(22)	& 3.7394(8) &	3.7147(10)&	3.6912(28)\\
20.40(24)	& 3.7292(8) &	3.7037(10)&	3.6807(27)\\
22.46(23)	& 3.7198(8) &	3.6940(10)&	3.6704(27)\\
24.64(30)	& 3.7093(9) &	3.6838(9)&	3.6612(29)\\
27.36(6)		& 3.6984(7) &	3.6733(9)&	3.6489(26)\\
28.68(16)	& 3.6928(7) &	3.6683(9)&	3.6428(25)\\
30.78(20)	& 3.6838(7) &	3.6606(9)&	3.6342(26)\\
32.79(22)	& 3.6767(7) &	3.6518(9)&	3.6252(24)\\
34.75(17)	& 3.6699(7) &	3.6437(9)&	3.6166(24)\\
36.53(14)	& 3.6623(7) &	3.6366(9)&	3.6102(24)\\
38.14(16)	& 3.6544(7) &	3.6285(9)&	3.6025(24)\\
40.00(14)	& 3.6484(7) &	3.6195(9)&	3.5978(24)\\
42.28(20)	& 3.6437(7) &	3.6128(8)&	3.5794(23)\\
44.19(6)		& 3.6364(7) &	3.6068(9)&	3.5724(23)\\
46.64(14)	& 3.6291(7) &	3.6003(8)&	3.5608(22)\\
49.40(14)	& 3.6198(7) &	3.5930(8)&	3.5487(23)\\
51.61(22)	& 3.6128(7) &	3.5866(8)&	3.5398(22)\\
54.54(21)	& 3.6049(7) &	3.5781(8)&	3.5249(22)\\
57.05(20)	& 3.5983(7) &	3.5713(8)&	3.5143(23)\\
\hline\hline
\end{tabular}
\caption{Lattice parameters as a function of pressure, given as pseudo-cubic lattices parameters related to the orthorhombic parameters $a_{\mathrm o}$, $b_{\mathrm o}$ and $c_{\mathrm o}$ by $a_{\mathrm{pc}} = a_{\mathrm o}/\sqrt{2}$, $b_{\mathrm{pc}} = b_{\mathrm o}/2$ and $c_{\mathrm{pc}} = c_{\mathrm o}/\sqrt{2}$. The pressure can be considered hydrostatic up to 40 GPa only (see text).}
\label{tab:latticeconstants}
\end{center}
\end{table}

\begin{figure}[tb]
\includegraphics[width=0.45\textwidth]{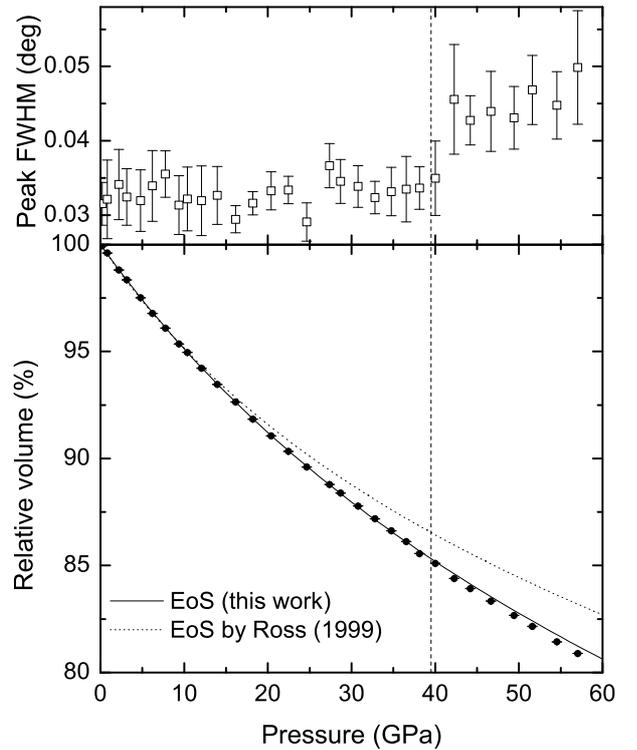}
\caption{(top) Width of Bragg peaks, averaged over selected reflections. (bottom) Evolution of the relative volume with pressure, together with the equation of state calculated by Ross \cite{Ross1999} (dotted line) and from this work (solid line). The vertical line marks the onset of non hydrostatic stress. Only data below this limit were used for the determination of the equation of state.}
\label{fig:volandstrain}
\end{figure}

We followed carefully the width of the diffraction peaks under pressure. Some peaks showed a marked broadening between 40 and 60 GPa, as shown in figure \ref{fig:volandstrain} (top), associated to a very slight anomaly in the evolution of the volume and lattice parameters. Alhough it is tempting to interpret this behaviour as the signature of a subtle phase change, not uncommon in perovskites, we note that this change is not associated to any change neither in the diffraction pattern (in terms of emergence/vanishing of diffraction lines), nor in the Raman spectra as will be detailed in the next section. After a comparison with a second data set recorded with neon as a pressure transmitting medium (not shown here), we attributed this change to the onset of non-hydrostatic stress, similar to the observations reported by Downs \etal{} on monoclinic aegirine \cite{Downs2006}. Details of this analysis and information that can be learned about the stress anisotropy in the cell will be detailed elsewhere. For the purpose of this work, we shall retain that (i) the $Pnma$ structure is stable up to 60 GPa and (ii) the pressure can be considered hydrostatic up to 40 GPa only for this experiment. 

The evolution of the volume with pressure at room temperature is reported in figure \ref{fig:volandstrain}. In the so-called $F$-$f$ plot \cite{Jeanloz1991} (not shown), our data between 0 and 40 GPa can be represented by a straight horizontal line, indicating that the data can be satisfactorily fitted by a second-order Birch-Murnaghan equation of state (EoS). The fit was performed by the program EoSFit \cite{Angel2001}. This fit yields a bulk mudulus of $K_0=181.0(6)$ GPa for a pseudo-cubic volume $V_0=55.830(17)$ \AA$^3$ and a constrained value of $K'_0=4$. The same procedure was used to determine the compressibilities of the three axes. Similarly, a second-order Birch-Murnaghan EoS was found sufficient in all cases. We find $K_{a} = 185.8(1.0)$ GPa, $K_{b} = 182.6(1.1)$ GPa and $K_{c} = 174.5(2.3)$ GPa.

\begin{table}[tb]
\begin{center}
\renewcommand{\arraystretch}{1.2}
\begin{tabular}{>{$}l<{$} m{1.7cm} m{1.7cm} m{1.7cm} m{1.7cm}}
\hline\hline
		& This work & Ref. \cite{Ross1999} 	& Ref. \cite{Xiong1986} & Ref. \cite{Wu2004} \\ 
$eq.$	& BM2			& BM3							& BM3							& M\\ 
K_0	& 181.0(6)	& 170.9(1.4)				& 210(7)						& 222(9)\\ 
K'_0	& \it 4		& 6.6(03)					& \it 5.6					& \it 4\\\hline
$eq.$	& BM2			& M							\\
K_{a}	& 185.8(1.0)& 168.3(1.9)\\
K'_{a}& \it 4		& 7.0(4)\\
K_{b}	& 182.6(1.1)& 175.3(1.5)\\
K'_{b}& \it 4		& 6.6(3)\\
K_{c}	& 174.5(2.3)& 168.7(2.1)\\
K'_{c}& \it 4		& 5.7(5)\\
\hline\hline
\end{tabular}
\caption{Parameters of the equation of state (EoS) from this work compared to the values determined by Ross and Angel \cite{Ross1999}, Xiong \etal{} \cite{Xiong1986} and Wu \etal{} \cite{Wu2004}. The EoS is indicated as follows: M = Murnaghan EoS, BM$i$ = $i^{\mathrm{th}}$-order Birch-Murnaghan EoS. Values in italic indicate parameters that are either unrefined or implied by the model used.}
\label{tab:eos}
\end{center}
\end{table}

In table \ref{tab:eos}, we compare our parameters of the EoS with previous results. The values reported by Ross and Angel \cite{Ross1999} were recorded on a single crystal up to 10 GPa. They have found $V_0=55.941(4)$ \AA$^3$, $K_0=170.9(14)$ GPa, and $K'=6.6(3)$ from the fit of the $P$-$V$ data with a third-order Birch-Murnaghan EoS. Although this result is in very good agreement with previous determination of $K_0$ by ultrasonic measurements (see Ref. \cite{Ross1999} and references therein), this equation of state extrapolated up to 40 GPa does not reproduce our experimental data at high pressures (figure \ref{fig:volandstrain}). The fit to our data yields different values with a bulk modulus $K_0$ exceeding by 10 GPa (6\%) the value given by Ross. This difference might be explained by the fact that they used $P$-$V$ data up to 10 GPa to constrain their EoS, whereas we used values up to 40 GPa. A fit of our data up to 10 GPa only yields $K_0 = 172(3)$ GPa and $K'_0 = 5.1(7)$, in agreement with the results by Ross. On the other hand, our values are far from Xiong \etal{} \cite{Xiong1986}, probably because of the relative inaccuracy of the lattice parameters derived from their experiment in non-hydrostatic conditions. The same limitations apply to the result by Wu \etal{} \cite{Wu2004}, obtained from a fit to a Murnaghan equation of state, which is known to be inappropriate for large compressions.

The values for the axial compressibilities are also reported in table \ref{tab:eos} and compared to the values by Ross and Angel. All values are slightly higher than their results, consistent with the higher bulk modulus. We find however that the least compressible of the three axes is the $a$ axis, and not the $b$ axis as reported in \cite{Ross1999}. This disagreement should be taken with care, given the fact that the degree of anisotropic compression remains very small. This may have an influence on the estimation of the pressure-induced distortion, as will be discussed later.

For the purpose of estimating the tilt angles from diffraction data, it is useful to recall how the tilt system in $Pnma$ perovskites can be described: rather than with Glazer's tilt system, two tilt angles $\phi$ and $\theta$ are often used, representing tilts around the $[010]$ and $[101]$ directions respectively, relative to the pseudo-cubic axes, as defined by Zhao \cite{Zhao1993,Zhao1993a,Mitchell2002}. Alternatively, the two symmetry-independent Ti-O1-Ti ($\alpha_1$) and Ti-O2-Ti ($\alpha_2$) angles can be used. Experimentally, tilt angles are best determined from atomic positions obtained by full structural refinements. Under the assumption of undistorted octahedra, these angles can be calculated from the lattice parameters only, using the following geometrical relations ($Pnma$ settings):
\begin{equation}
\begin{array}{l l l}
\cos\phi & = & c\sqrt{2}/b\\
\cos\theta & = & c/a
\end{array}
\label{eq:formulaangles}
\end{equation}
The $\alpha_i$ angles can in turn be calculated from \cite{Mitchell2002}:
\begin{equation}
\begin{array}{l l l}
\cos\theta & = & \cos\{(180-\alpha_1)/2\}\\
\cos\phi & = & \cos\{(180-\alpha_2)/2\}/\sqrt{\cos\theta}
\end{array} 
\label{eq:1}
\end{equation}
With this parametrization, an increase of the distortion corresponds to a decrease of the angles $\alpha_i$. At ambient conditions, the angles calculated from the atomic positions found in Ref. \cite{Sasaki1987} are $\phi=8.9$\degre, $\theta=11.6$\degre, $\alpha_1 = 156.8$\degre and $\alpha_2 = 155.8$\degre.

\begin{figure}[tb]
\includegraphics[width=0.45\textwidth]{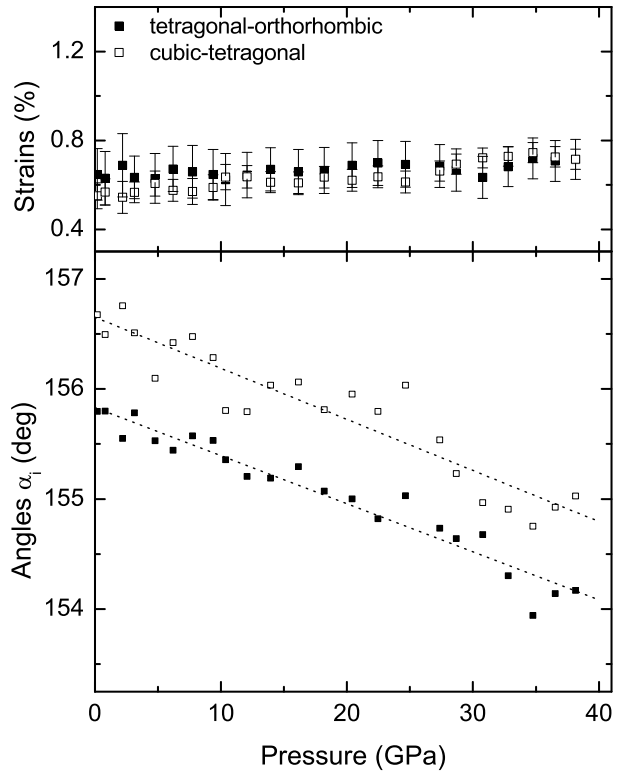}
\caption{(top) Pressure dependence of spontaneous strains under pressure. (bottom) Evolution of the $\alpha_i$ angles under pressure, as calculated from equation (\ref{eq:1}). The dotted line is a linear fit to the data.}
\label{fig:strainsandangles}
\end{figure}

We calculated the tilt angles from the experimental lattice constants using the geometrical equations (\ref{eq:1}). Those relations are known to underestimate the actual tilt angles determined from atomic positions \cite{Mitchell2002}, so we additionally rescaled the values so as to match the angles at room pressure. The evolution is shown in figure \ref{fig:strainsandangles} (bottom). The angles decrease slightly with pressure, with an average rate of -0.045\degre/GPa. 

It is also possible to quantify the distortion in $Pnma$ perovskites by spontaneous strains. As pointed out in Ref.  \cite{Ross1999,Redfern1996,Carpenter2001,Mitchell2002}, spontaneous strains of interest for the orthorhombic structure can be calculated from the pseudo-cubic lattice parameters as follows: $\eps_{\mathrm{ct}} = |c-a_0|/a_0$ measures the distortion from cubic to tetragonal and $\eps_{\mathrm{ot}} = |a-b|/a_0$ the distortion from orthorhombic to tetragonal, were $a_0$ is the cubic lattice constant approximated here as $V^{1/3}$. These quantities are plotted in figure \ref{fig:strainsandangles} (top). Both strains increase up to 40 GPa, but so slightly that it almost remains constant within the experimental uncertainty. This small variation is consistent with the observation by Ross and Angel \cite{Ross1999} who had observed no pressure-dependence of these strains between 0 and 10 GPa, and confirms that these strains are not well adapted to the study of the very weak pressure-induced distortion in CTO.

\section{Raman spectroscopy}

The group theoretical analysis for CTO in the $Pnma$ space group yields 24 Raman active phonon modes decomposed as $7A_g + 5B_{1g} + 7B_{2g} + 5B_{3g}$. The Raman spectrum of CTO has been studied experimentally by several authors in the past at ambient conditions, as a function of temperature \cite{Balachandran1982,McMillan1988,Zelezny2002}, under pressure up to 26 GPa \cite{Gillet1993} and under high-pressure and temperature \cite{Gillet1993a}. In spite of these works, a comprehensive assignment of the 24 Raman active phonons is still lacking. This is due to (i) band overlapping of the numerous Raman modes, (ii) the small distortion from the cubic perovskite which may result in narrow splittings and low intensities for some Raman modes and (iii) an intense background scattering. This background can be described as two broad bands in the 200-550 and 600-750 cm$^{-1}$ ranges, usually accounted for by second-order scattering processes, by analogy with the cubic perovskite SrTiO$_3$ in which a very similar background is observed although first-order Raman modes are forbidden by symmetry. A partial assignment of the 7 $A_g$ and some $B_{2g}$ modes, recalled in table \ref{tab:ramanmodes}, was proposed by Mc Millan \etal{} \cite{McMillan1988}. The frequencies and symmetries of Raman active phonon modes were also predicted by ab-initio calculations \cite{Cockayne2000} but the results appear to be strongly dependent on the choices made on the structure relaxation (atomic positions and/or lattice parameters) in the calculation, and has not been used so far to clarify the assignment. We also note that both studies contain some confusion in the group-theoretical analyses due to the choice of the orthorhombic space group setting $Pnma$ or $Pbnm$. A correspondence between the two settings and the details of the Raman tensors in both cases can be found for example in \cite{Smirnova1999}. The $Pnma$ setting is used throughout this paper.

\begin{figure}[tb]
\includegraphics[width=0.45\textwidth]{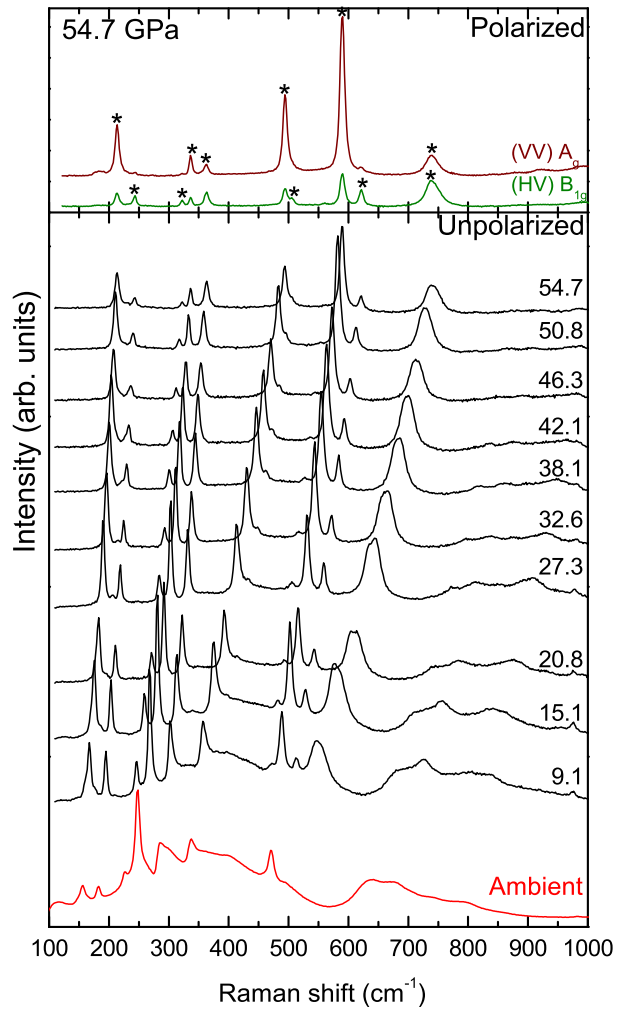}
\caption{(top) Polarized spectra recorded at 54.7 GPa in parallel (VV) and crossed (HV) polarization. The modes identified as $A_g$ (resp. $B_{1g}$) are marked with stars $^*$. (bottom) Selected unpolarized spectra recorded under pressure. The spectrum at ambient conditions was recorded on a different crystal.}
\label{fig:ramanspectra}
\end{figure}

\begin{figure}[tb]
\includegraphics[width=0.45\textwidth]{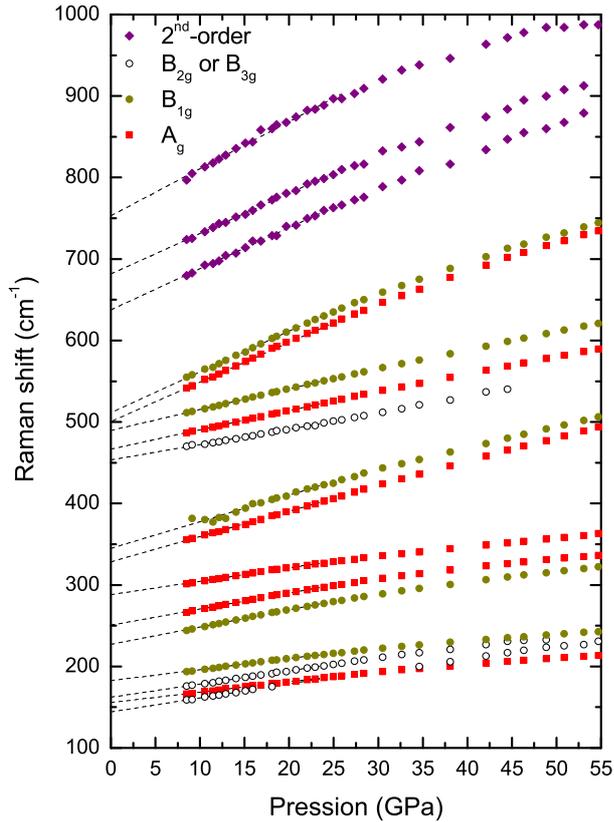}
\caption{Evolution of the Raman shifts with pressure. The strongest peaks are marked with full symbols while the open symbols indicate weaker bands. The purple diamonds correspond to second-order features. The dashed line are linear fits to the data in the 0-25 GPa range.}
\label{fig:ramanshifts}
\end{figure}

Selected Raman spectra recorded under pressure are presented in figure \ref{fig:ramanspectra}. The intensity of the broad bands associated with second-order processes gradually decreases with pressure. At 30 GPa, the band originally situated in the 200-550 cm$^{-1}$ range can be considered to have completely vanished, while some traces of the second band are still visible in the high-frequency part of the spectrum at 800-1000 cm$^{-1}$. This phenomenon is reminiscent of the evolution of the Raman spectrum of SrTiO$_3$ under pressure and probably has the same origin \cite{Grzechnik1997,Guennou2010}. A total of 12 sharp first-order peaks are identified, as well as a broader band at higher frequency that was fitted with two peaks, although more contributions might in fact be overlapping. The evolution of the Raman shifts for the identified bands is shown in figure \ref{fig:ramanshifts}. All modes harden under pressure, although with somewhat different slopes and curvatures. We determined the initial slopes of the different modes by linear fits in the 0-25 GPa range (table \ref{tab:ramanmodes}). The linewidths (not shown) vary little with pressure and do not show any change that would indicate a phase transition. We measured the spectrum after the experiment (after opening of the cell) and ensured that the typical spectrum of CTO was recovered. 

Polarized spectra were recorded at 55 GPa, the highest pressure investigated (figure \ref{fig:ramanspectra}). The orientation of the crystal used in the Raman experiment was previously determined from x-ray diffraction in the DAC at room conditions. In this geometry the incident laser propagation direction lies very close to the $z$-axis (less than 10\degre off) so that the polarization of the laser lies in the $(xy)$ plane, allowing us to separate $A_g$ and $B_{1g}$ modes in parallel (VV) and crossed (HV) polarizations respectively. The spectra in figure \ref{fig:ramanspectra} were obtained after rotating the DAC so as to maximize the contrast between the two configurations, in order to be as close as possible to the crystallographic axes. Our polarized spectra therefore allow us to unambiguously assign the modes at 183, 227, 344 and 489 cm$^{-1}$ to $B_{1g}$ symmetry, all of them being strong in the (HV) spectrum and almost absent in the (VV) spectrum (all frequencies are refered to at room conditions). Similarly, the modes at 155, 251, 328 and 467 cm$^{-1}$ can be assigned to $A_g$ symmetry. Those four peaks are also present in the (HV) spectrum, probably as a result of polarization leakage and imperfect sample orientation. The situation is less straightforward for the mode at 288 and the intense modes at 500-510 cm$^{-1}$ which are observed in both geometries with almost the same intensity. We assign the mode at 288 cm$^{-1}$ to $A_g$ symmetry, although its rather strong intensity in the (HV) spectrum might indicate a mode superposition. The modes at 500-510 cm$^{-1}$ are peculiar in so far as they are hardly visible at room conditions but gain intensity under pressure. These modes correspond to the strong broad band at 493 cm$^{-1}$ in the $B_{1g}$ spectrum noted by Mc Millan and Ross \cite{McMillan1988}, who mentioned that it could be due to either first- or second-order scattering. We believe that a second-order nature of these two modes can be ruled out. First, they exhibit a high intensity at both high pressure and low temperature \cite{Zelezny2002}. Second, they are observed in both polarization geometries (VV) and (VH) thus they cannot be pure overtones. We therefore explain them by a superposition of modes with $A_g$ and $B_{1g}$ symmetry. In addition, three very weak bands are observed at 144, 162 and 453 cm$^{-1}$. Their weak intensity does not enable their symmetry to be determined from the polarized spectra. It is nonetheless possible to obtain some information from the fact that the mode at 144 cm$^{-1}$ crosses the mode at 155, which was assigned $A_g$ symmetry. Since two modes crossing each other have to be of different symmetries, the mode at 144 cm$^{-1}$ is expected to be $B_{ig}$. Similarly, the weak mode at 162 cm$^{-1}$ crosses the mode at 183 cm$^{-1}$. This suggests $B_{2g}$ or $B_{3g}$ symmetries for these weak modes, only observed because of imperfections in the sample orientation and polarization conditions. 

These assignments are summarized in table \ref{tab:ramanmodes} and compared to the previous assignment by Mc Millan and Ross \cite{McMillan1988}. We do not confirm the assignment of the modes at 183 and 227 cm$^{-1}$ to $A_g$ symmetry as they proposed, but we note that they have not observed these two modes in our configuration ($z(xx)z$ or $z(yy)z$). 

In a second step, it is useful to associate to each Raman mode specific atomic displacements. A significant amount of work has been carried out in the past for $Pnma$ perovskites  and vibrational patterns of the Raman active phonon modes have been proposed by Iliev \etal{} \cite{Iliev1998,Abrashev2002} in their study of various AMnO$_3$ perovskites. The numbering of the Raman modes in the following is taken from these studies. This work requires the knowledge of the phonon eigenvectors, which can be obtained by either ab-initio or empirical lattice dynamics calculations, often carried out using a shell model. As far as CTO is concerned, several ab-initio \cite{Parlinski2001,Cockayne2000,Lebedev2009} and shell model calculations \cite{Gillet1993,Calleja2003} have been reported. Unfortunately, these calculations have been performed for other purposes than mode assignment and show too poor an agreement between calculated and experimental Raman frequencies to be used for our study. Information can nonetheless be gained from comparison with Raman data and simulations of isostructural CaBO$_3$ perovskites. Several such perovksites have been studied in the past for B = Ge, Sn, Zr, Mn  \cite{Durben1991,Gupta1999,Abrashev2002,Tarrida2009}. All of them have the $Pnma$ orthorhombic structure and the distortion from the cubic perovskite in CTO, as measured for example by the tilt angles, is intermediate between CaGeO$_3$ and CaMnO$_3$. Since the B cation sits on a center of symmetry of the crystal, it does not participate in any Raman active mode, and can be expected to have a moderate influence on the Raman frequencies. This is supported by the Raman work by Tarrida \etal{} \cite{Tarrida2009} across the solid solution CaZrO$_3$-CaSnO$_3$, which shows that most observed Raman frequencies are only weakly dependent on the B cation. Conversely, their analysis of the Raman spectra across the CaZrO$_3$-SrZrO$_3$ system clearly points out the Raman modes that strongly depends on the A cation. The comparison for the different CaBO$_3$ perovskites is shown in table \ref{tab:ramanmodes2}. The agreement between the experimental $A_g$ frequencies is remarkable. The agreement is much more difficult to verify for the $B_{ig}$ modes due to the incomplete mode identification. It is nonetheless possible to proceed to an identification of the pattern of vibrations from the CaMnO$_3$ \cite{Abrashev2002}. The calculated Raman frequencies for CaMnO$_3$, on which the mode assignment is based, is also recalled. For comparison, the results of the shell model calculations for CaMnO$_3$ by Sopracase \etal{} \cite{Sopracase2010} are also indicated. Their model was successfully fitted against the infrared active frequencies, but shows marked differences to the Raman frequencies.

Let us consider first the set of modes that involve displacements of the Ca cation in the $xz$ plane or along the $y$ direction. The three $A_g$ modes $A_g(5,6,7)$ can be identified in all the compounds at very similar frequencies that fit well with the shell model calculations by Abrashev \etal{} on CaMnO$_3$ \cite{Abrashev2002}. Among the $B_{ig}$ modes involving Ca-displacements, few comparisons can be made, except for $B_{2g}(7)$ that has been observed in CaGeO$_3$ and CaZrO$_3$. This attribution is very well corroborated by the study of (Ca,Sr)ZrO$_3$ system, which shows that these four modes are by far the most A-cation dependent \cite{Tarrida2009}. In CaTiO$_3$ the corresponding mode frequencies are 251, 288 and 328 cm$^{-1}$. In addition, the associated mode at 344 cm$^{-1}$ matches very well the $B_{1g}(5)$ mode that also implies Ca shifts along the $y$ direction, which supports our symmetry-based assignment. 

We now turn to the highest frequency modes that involve B-O bond stretching. Three such modes are expected in the 500-550 cm$^{-1}$ range. The $A_g(1)$ mode was observed in CaTiO$_3$, CaZrO$_3$ and SrZrO$_3$ and it was shown \cite{Tarrida2009} that its frequency was almost insensitive to the $A$ cation. This mode is closely related to the $B_{1g}(2)$ mode that involves the same movements. This supports our assignment of a superposition of $A_g+B_{1g}$ for the bands at 505 cm$^{-1}$. Other modes involving bond stretching are expected from the calculation at higher frequencies ($B_{1g}(1), B_{2g}(1), B_{3g}(1)$) but have not been observed conclusively in any of the compounds.

Of particular interest are the modes corresponding to rotations of the octahedra around the $[010]$ and $[101]$ directions respectively: $A_g(2)$ and $A_g(4)$. A comparison with the modes in CaMnO$_3$ suggests they correspond to the two low-frequency $A_g$ modes. One of them in CTO can be identifed as the mode at 155 cm$^{-1}$. An ambiguity remains for the second rotational mode: the comparison with the other compounds suggests this could be the mode at 188 cm$^{-1}$ identified as an $A_g$ mode by McMillan and Ross \cite{McMillan1988}, even though we could not confirm this assignment. The two low-frequency $B_{1g}(3)$ and $B_{1g}(4)$ modes also involve rotations of the octahedra and our values for CTO compare well with the frequency measured in CaGeO$_3$ and CaMnO$_3$.

\begin{table}[tb]
\begin{center}
\begin{tabular}{c c c c}
\hline\hline
$\omega_0$ 	& \multicolumn{2}{c}{Symmetry} 			& Slope 			\\
(cm$^{-1}$) & This work & Ref. \cite{McMillan1988} & (cm$^{-1}$/GPa) \\
144 & $B_{2g}$ or $B_{3g}$	& 								& 1.72	\\
155 & $A_g$				 	& $A_g$							& 1.28	\\
162 & $B_{2g}$ or $B_{3g}$	& 								& 1.62	\\
183 & $B_{1g}$				& $A_g (+ B_{2g}?)$			& 1.35	\\
227 & $B_{1g}$				& $A_g (+ B_{2g}?)$			& 2.12	\\
251 & $A_g$				 	& $A_g (+ B_{2g}?)$			& 1.97	\\
288 & $A_g$				 	& $A_g$							& 1.64	\\
328 & $A_g$				 	& $A_g$							& 3.11	\\
344 & $B_{1g}$				& 									& 3.27	\\
453 & $B_{2g}$ or $B_{3g}$	& 								& 1.89	\\
467 & $A_g$				 	& $A_g$							& 2.36	\\
489 & $B_{1g}$				& 									& 2.56	\\
500 &	$A_g$					&									& 4.84	\\
510 &	$B_{1g}$				&									& 4.96	\\
\hline
\multicolumn{4}{c}{Second-order bands}\\
638 &							&									& 5.04\\
682 &							&									& 4.95\\
753 &							&									& 5.78\\
\hline\hline
\end{tabular}
\caption{Summary of the observed Raman modes.}
\label{tab:ramanmodes}
\end{center}
\end{table}

\begin{table*}[tb]
\begin{center}
\renewcommand{\arraystretch}{1.3}
\begin{tabular}{l c c c c c c c r}
\hline\hline
	& CaGeO$_3$ & CaTiO$_3$ & \multicolumn{3}{c}{CaMnO$_3$} 	& CaSnO$_3$ & CaZrO$_3$ \\
\cline{4-6}
Mode	& Ref. \cite{Durben1991} & This work	& \multicolumn{2}{c}{Ref. \cite{Abrashev2002}} & Ref. \cite{Sopracase2010}	& Ref. \cite{Tarrida2009} & Ref. \cite{Orera1998}	& \\
\cline{4-5}
symmetry		& Exp.		& Exp.		& Exp. 	& Calc.	& Calc.	& Exp.	& Exp. & Mode description \\
\hline
$A_g(2)$		& 151			& 155 		& 160		& 154 &	146 & 145 		& 145 & In-phase $y$-rotation\\
$A_g(4)$		& 171			&  -			& 184		& 200 &	229 &  - 		&  -	& In-phase $x$-rotation\\
$A_g(7)$		& 265			& 251			& 243		& 242 &	323 & 265 		& 263 & Ca and O1, $x$-shifts\\
$A_g(5)$ 	& 284			& 288			& 278		& 299 &	336 & 277 		& 287 & Ca, $z$-shifts\\
$A_g(6)$		& 328			& 328			& 322		& 345 &	406 & 355 		& 358 & Ca and O1, $x$-shifts\\
$A_g(3)$		&  -			& 467			& 487		& 467 &	529 & 442 		& 439 & Out-of-phase B-O bending\\
$A_g(1)$		&  -			& 500			& 	-		& 555 &	657 &  - 		& 543	& In-phase B-O stretching\\\hline
$B_{1g}(3)$	& 184			& 183			& 179		& 178 &	207 &  - 		& - 	& Out-of-phase $y$-rotation \\
$B_{1g}(4)$	& 237			& 227			& 	-		& 281 &	265 & - 			& 227 & In-phase $x$-rotation		\\
$B_{1g}(5)$	& 358			& 344			& 	-		& 354 &	450 & - 			& 305 & Ca and O1, $y$-shift\\
$B_{1g}(2)$	&  -			& 510			& 	-		& 536 &	656 & - 			& 439 & Out-of-phase B-O stretching \\
$B_{1g}(1)$	&  -			& 489			& 	-		& 743 &	700 & -			& 547 & Out-of-phase B-O stretching\\\hline
$B_{2g}(4)$	& 162			& 	-			& 	-		& 148 &	229 & -			& 190 & Out-of-phase $z$-rotation		\\
$B_{2g}(7)$	& 248			& 	-			& 	-		& 232 &	249 & -			& 212 & Ca and O1, $z$-shifts\\
$B_{2g}(5)$	& 	-			& 	-			&  -		& 292 &	306 & -			& 234 & Ca and O1, $z$-shifts		\\
$B_{2g}(6)$	& 	-			& 	-			&  -		& 366 &	420 & -			& 418 & Ca, $x$-shifts		\\
$B_{2g}(3)$	& 	-			& 	-			&  -		& 453 &	503 & -			& 469 & Out-of-phase B-O bending \\
$B_{2g}(2)$	& 	-			& 	-			&  -		& 485 &	541 & -			& 543 & In-phase B-O bending		\\
$B_{2g}(1)$	& 	-			& 	-			&  -		& 749 &	713 & - 			& - 	& In-phase B-O stretching	\\\hline
$B_{3g}(5)$	& 	-			& 	-			&  -		& 290 &	234 & - 			& - 	& Ca, $y$-shifts	\\
$B_{3g}(4)$	&  -			& 	-			& 320		& 304 &	365 & - 			& - 	& In-phase $z$-rotation	\\
$B_{3g}(3)$	& 380			& 	-			& 	-		& 459 &	503 & - 			& - 	& Out-of-phase B-O bending	\\
$B_{3g}(2)$	& 496			& 	-			& 564		& 541 &	655 & - 			& - 	& Out-of-phase B-O stretching \\
$B_{3g}(1)$	& 	-			& 	-			&  -		& 754 &	719 & - 			& - 	& Out-of-phase B-O stretching \\
\hline\hline
\end{tabular}
\caption{Comparison of the observed Raman modes for different Ca$B$O$_3$ perovskites, ordered roughly from the less distorted to the most distorted structure. The numbering and description of the Raman modes follows Ref. \cite{Abrashev2002}. The assignment of many $B_{ig}$ Raman modes is unclear due to incomplete information. See the text and references for details. }
\label{tab:ramanmodes2}
\end{center}
\end{table*}

\section{Discussion}

\subsection{Stability of the $Pnma$ structure}

Our Raman and X-ray scattering experiments show that the $Pnma$ orthorhombic structure of CTO remains stable up to 60 GPa. This result is in contradiction with the report by Xiong \etal{} \cite{Xiong1986} of a phase transition to a hexagonal structure at 10 GPa, but confirms and extends the study by Gillet \etal{} \cite{Gillet1993} whose Raman scattering experiment had not revealed any phase transition up to 26 GPa. The change interpreted by Xiong \etal{} \cite{Xiong1986} as a phase transition was most probably due to the solidification of the ethanol-methanol pressure transmitting medium around 10 GPa, all the more that they could not reproduce this transition in a subsequent experiment carried out without any transmitting medium. Moreover, the absence of a phase change does not confirm the ab-initio calculation by Wu \etal{} \cite{Wu2005a} who had predicted a transition to the post-perovskite $Cmcm$ phase at 30 GPa \cite{Wu2005a}. This disagreement could be partially due to the ab-initio calculations being performed at 0 K while our measurements are performed at room temperature and do no exclude the possibility of a phase transition at low temperatures. Alternatively, it could be that high temperatures are necessary to break the bonds involved in the reconstructive perovskite to post-perovskite transition \cite{Wu2005a}.

\subsection{Evolution of the distortion under pressure}

We now want to discuss the evolution of the distortion with pressure, especially in the high-pressure regime. First, we shall recall briefly the main features of the valence bond sum model by Zhao \cite{Zhao2006}. For a given site $i=A$ or $B$, the valence bond sum $V_i$ is
\begin{equation}
V_i = \sum_j^N\exp\left(\frac{R_0-R_{ij}}{b}\right)
\end{equation}
where $N$ is the coordination number, $R_{ij}$ is the bond length, $R_0$ a constant specific to an anion-cation pair and $b = 0.37$ \AA{} a universal constant. The valence bond sum equals the formal valence of the cation at ambient conditions. It is then assumed that the pressure-induced changes in the valence bond sum for sites A and B are equal ("bond valence sum matching principle"). It can then be shown that the compressibility ratio $\beta_B/\beta_A$ is estimated by $M_A/M_B$, where $M_i$ is the site parameter defined from the average bond lengths $R_i$ by
\begin{equation}
M_i = \frac{R_iN}{b}\exp\left(\frac{R_0-R_{i}}{b}\right).
\end{equation}
In addition, it has been shown later \cite{Zhao2006} that the same model could be used to calculate the evolution of the tilt angle under pressure from the knowledge of the detailed structure at ambient conditions and the pressure-dependence of the lattice parameters. For $A^{2+}B^{4+}$O$_3$ perovskites such as CaTiO$_3$, this model yields $\beta_B/\beta_A < 1$, i.e. the distortion should increase under pressure, and the calculated increasing rate of the tilt angle is -0.131\degre/GPa, in good agreement with experimental data collected up to 10 GPa \cite{Zhao2004,Zhao2006}. In the following, we will analyze the evolution of the distortion from our data, and check this model against the evolution up to the highest pressures measured.

Within the model by Zhao, the angles $\alpha_i$ are determined from the linear compressibilities of the lattice parameters $\beta_a$, $\beta_b$, $\beta_c$ and the Ti-O bonds, using the relation
\begin{equation}
\alpha_i = 2 \sin^{-1}[\exp(\Delta\beta_i P)\sin(\alpha_{0,i}/2)]
\label{eq:anglezhao}
\end{equation}
where $\Delta\beta_1 = \beta_{B-O1} - \beta_b$ and $\Delta\beta_2 = \beta_{B-O2} - (a^2\beta_a + c^2\beta_c)/(a^2+c^2)$. The average bond compressibility $\beta_B$ is a good estimate for $\beta_{B-Oi}$. $\beta_B$ in turn is calculated from the variation of the valence bond sum $V_B$ under pressure:
\begin{equation}
\beta_B = \frac{b}{R_{0}}\frac{\mathrm d\ln V_B}{\mathrm d P}
\end{equation}
Finally, $V_B$ can be estimated at a given pressure in the following way: we calculate for each pressure the average of the bond valence sums for sites $A$ and $B$ using the "fixed-coordinates" model, whereby the atomic coordinates in the cell are kept equal to their values at ambient conditions, and only the lattice parameters are set to their experimental values. It was shown in \cite{Zhao2006} that this yields a good estimate for the pressure-induced change in the bond valence sum $\Delta V_B$.

\begin{figure}[tb]
\includegraphics[width=0.45\textwidth]{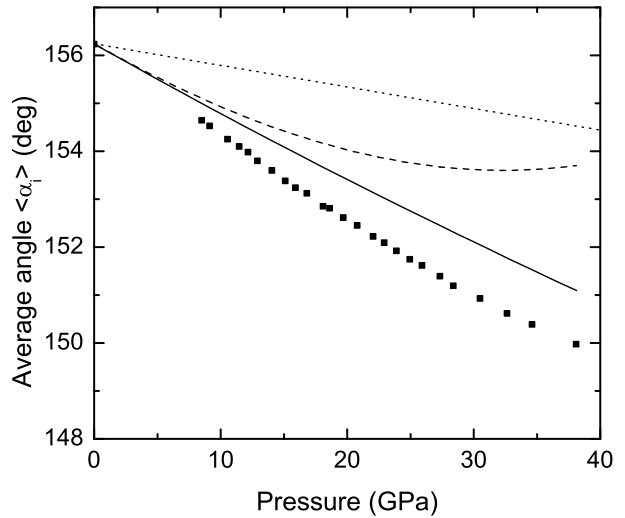}
\caption{Evolution of the average Ti-O-Ti angle under pressure calculated in different ways: calculated from the lattice constants using the geometrical relations (\ref{eq:1}) (dotted line), calculated from the bond valence sum model with pressure-independent compressibilities (solid line), calculated from the valence bond sum model with pressure-dependent compressibilities (dashed line), calculated from the frequency of the $A_g(2)$ Raman mode (symbols).}
\label{fig:angles}
\end{figure}

For the calculation of the tilt angles, we take the initial structure from Sasaki \cite{Sasaki1987} and the values for $R_0$ (1.815 \AA{} for Ti-O and 1.967 \AA{} for Ca-O) are taken from Ref. \cite{Zhao2004}. For the calculation of the compressibilities, we proceeded in two steps. In a first step, for comparison purposes with the results by Zhao \cite{Zhao2006}, we estimated the linear compressibility of axes  as well as $\mathrm d\ln V_B/\mathrm dP$ by a linear fit to our data between 0 and 10 GPa. With pressure-independent compressibilities calculated in this way, we find an average rate of $\mathrm d\langle\alpha_i\rangle/\mathrm dP=-0.134$ \degre/GPa, in excellent agreement with the -0.131\degre/GPa by Zhao \cite{Zhao2006}, showing that the differences found in the previous section regarding the axis compressibilities are small enough to be neglected for the calculation of the tilt angles. The result of this first calculation is shown in figure \ref{fig:angles} (solid line). The comparison with the result found from the direct calculation of the tilt angles from the lattice constants alone (dotted line in figure 6) shows that the geometrical relations (\ref{eq:1}) do not yield a correct angle change rate (-0.045 vs. -0.134\degre/GPa). This result most probably reflects that the hypothesis of undistorted octahedra is invalid at high pressures. 

In a second step, we tentatively performed the same calculation over the full pressure range with pressure-dependent axial compressibilities. Of course, we expect the model to break down at some point, since the parameters $b$ and $R_0$ used to calculate the bond valence sum are determined by the requirement that the valence bond sum has to be equal to the formal charge of the cation at ambient pressure. The result is shown in figure \ref{fig:angles} (dashed line). The angles follow closely the results of the first calculation close to ambient pressure but show a much more pronounced curvature and even start to increase again around 35 GPa. This is not unrealistic, since a change in the compression mechanism could in principle result in such an evolution, as is presumably the case in ScZrO$_3$\cite{Andrault1991}. In CTO, this behavior remains to be verified experimentally, which would require full structural refinements to calculate tilt angles as accurately as possible. Unfortunately, full structural refinements on single crystals in this pressure range are technically challenging, and X-ray diffraction is not a very sensitive tool for the determination of oxygen positions. 

In this context, Raman spectroscopy may be an interesting alternative to follow the two compression mechanisms: bond compression and octahedra tilting. In principle, it is possible to estimate directly the compressibility of a polyhedron from high-pressure Raman measurements provided that a phonon mode can be found that involves in-phase expansion of all its bonds \cite{Lucazeau2003}. The relevant quantity is the mode compressibility that is proportional to $(\mathrm d\ln\omega/\mathrm dP)$. By comparing the relative increase of the chosen phonon frequencies, it is possible to estimate a compressibility ratio and decide whether the distortion increases or decreases under pressure. However, in a crystal with distorted polyhedra where most phonon modes involve mixed displacement of atoms, it is not straightforward that a suitable phonon mode exists that faithfully reflects the compression of the bonds or the polyhedra. As far as tilt angles are concerned, Iliev \etal{} have shown in their studies of a series of rare-earth manganates that the two rotational $A_g(2)$ and $A_g(4)$ mode frequencies scale with the tilt angles with the empirical rate 23.5 cm$^{-1}$/\degre \cite{Iliev2006}. This rule has also been found approximately valid in rare-earth scandates \cite{ChaixArxiv}. Unfortunately, $A^{2+}$Ti$^{4+}$O$_3$ perosvkites with the $Pnma$ structure are rare and the same systematic work can hardly be done. We may however expect a similar rule to hold true for CTO, i.e. the frequencies of the rotational modes to scale with the tilt angles. 

From our mode assignment, we oberve that the most pressure-dependent modes are the modes involving Ti-O bond stretching and Ca-shifts. On the other hand, the rotational modes at low frequencies show a much more moderate frequency increase under pressure. This is qualitatively consistent with the general picture of a volume reduction obtained by an overall compression of the bonds with only a slight variation of the tilt angles. The bands with the highest mode compressibility are the bands at 500-510 cm$^{-1}$ which we have assigned to Ti-O stretching. According to the vibrational pattern by Iliev, the mode best suited to the examination of the compression of TiO$_6$ octahadra would be the $B_{3g}(1)$ at high frequencies (octahedra "breathing" mode). This mode however, was not identified in CaTiO$_3$ nor in any of the Ca$B$O$_3$ compounds mentioned above. Moreover, the mode compressibility calculated from the evolution of bands at 500-510 cm$^{-1}$ is the strongest of all observed modes, and we cannot expect to extract compressibility ratios in agreement with the value now well established from crystallography. This reflects the need for a more precise identification of the relevant breathing modes.

Finally, we consider the evolution of the rotational modes and assume that the $A_g(2)$ and $A_g(4)$ scale with the tilt angles $\phi$ and $\theta$ respectively. In CTO, we have $\phi = 8.9$\degre and $\theta = 11.6$\degre. The $A_g(2)$ rotation mode was identified as the mode at 155 cm$^{-1}$ and we additionally assume the $A_g(4)$ is indeed the mode at 181, although it is emphasized that some ambiguity remains. From the evolution of their frequencies under pressure (1.28 and 1.35 cm$^{-1}$/GPa respectively), we calculate an average angle variation of -0.160\degre/GPa, which compares reasonably well to the -0.134\degre/GPa calculated by the valence bond sum model. It is therefore not unreasonable to consider that these phonon frequencies provide a good approximation of the average tilt angle. The average angle calculated in this way (and shifted to match the value at zero pressure) is plotted in figure \ref{fig:angles} together with the results of the different calculations. The evolution suggests that the increase of the tilt angles slows down as pressure increases, which is consistent with the prediction of the pressure-dependent valence bond sum model, although the pressure ranges involved are very different. Further work and more conclusive mode assignments are needed for a validation of this approach, but we believe that Raman spectroscopy can be a valuable tool to estimate tilt angles in the high-pressure region.

\section{Conclusion}

We have performed X-ray diffraction and Raman spectroscopy on single crystal CaTiO$_3$ in a diamond-anvil cell up to 60 GPa and 55 GPa respectively. The orthorhombic $Pnma$ structure remains stable over the whole pressure range investigated, which contradicts the ab-initio prediction of a transition to a post-perovskite phase at 30 GPa. We calculated the bulk modulus of CTO $K_0$ by a fit to the compression data in the 0-40 GPa range and found a value of 181.0(6), in good agreement with previous results. In addition, polarized Raman spectra were recorded at high pressure and have enabled us to clarify the Raman mode assignment of CTO, and identify reliably Raman modes associated to Ti-O bond stretching, Ca shifts and octahedra rotations. The analysis of the rotational Raman modes as well as calculations using the valence bond sum model both suggest that the weak distortion of the $Pnma$ structure increases under pressure but with a decreasing rate.

\section{Aknowledgments}

The authors are grateful for precious help from the ESRF staff, especially M. Mezouar at the ID27 Beamline for the allocation of inhouse beamtime, as well as J. Jacobs from the sample environment pool. Support from the French National Research Agency (ANR Blanc PROPER) is acknowledged.

\bibliographystyle{aip}
\bibliography{biblio}

\end{document}